\documentclass[reprint,superscriptaddress,aps,prl,]{revtex4-2}
\usepackage{amsmath}
\usepackage{amssymb}
\usepackage{graphicx}
\usepackage{epstopdf}
\usepackage[colorlinks=true]{hyperref}
\usepackage{physics}
\usepackage{mathrsfs}
\usepackage{comment}

\renewcommand\({\begin{equation}}	
\renewcommand\){\end{equation}}
\renewcommand\[{\begin{eqnarray}}	
\renewcommand\]{\end{eqnarray}}
\newcommand{\al}[1]{\begin{aligned}#1\end{aligned}}
\usepackage[dvipsnames]{xcolor}

\begin{document}

\title{Topological Spin Textures Enabling Quantum Transmission}

\author{Ji Zou}
\affiliation{Department of Physics, University of Basel, Klingelbergstrasse 82, 4056 Basel, Switzerland}
\author{Stefano Bosco}
\affiliation{QuTech, Delft University of Technology, Lorentzweg 1, 2628 CJ Delft, The Netherlands}
\author{Jelena Klinovaja}
\author{Daniel Loss}
\affiliation{Department of Physics, University of Basel, Klingelbergstrasse 82, 4056 Basel, Switzerland}

\begin{abstract}
Quantum spintronics is an emerging field focused on developing novel applications by utilizing the quantum coherence of magnetic systems. {A key challenge in this context is achieving scalable long-range quantum information transmission in magnetic systems.  Here, we propose a novel transmission scheme based on topological spin textures in a hybrid architecture combining a magnetic racetrack and localized spin qubits.} We demonstrate this principle by employing the domain wall (DW)|the most fundamental  texture|to   transport quantum signal  between distant  qubits. {We introduce a measurement-free protocol that utilizes DW mobility to enable high-fidelity and tunable entanglement generation.    Furthermore, we demonstrate that spin qubits can function as quantum stations on the racetrack, enabling flexible state transfer among fast-moving DWs on a single track. Finally, we discuss concrete material platforms to implement the proposed scheme. Our work introduces a new hybrid quantum platform that merges topological spin textures with solid-state qubits, offering a scalable architecture for quantum information processing and opening promising directions for quantum spintronics.} 
\end{abstract}
\date{\today}
\maketitle


{Topological spin textures, such as domain walls (DWs) and skyrmions, have long been explored for classical information transmission in spintronics, with landmark proposals like racetrack memory driving significant experimental advances in the control and manipulation of these textures~\cite{Parkin190,Fert:2013aa}. In recent years, the field has experienced rapid expansion with growing interest in quantum regimes~\cite{YUAN20221}. Significant efforts have been made  in the community both theoretically and experimentally,}  including the detection of quantum magnonic states~\cite{lachance2020entanglement,Kamra2023PRL}, their integration with NV centers~\cite{fukami2024magnon,zou2022prb}  and superconducting qubits~\cite{tabuchi2015coherent,PhysRevLett.129.037205}, and  the use of  topological textures as qubits for quantum devices~\cite{christina_prl_2021,xia2023universal,PhysRevLett.132.193601,Zou2023prr,qu2025density,PhysRevB.97.064401}.
However, a fundamental aspect of practical quantum spintronic devices|the reliable transmission of {quantum} information in magnetic systems|still remains elusive.  This has, in fact, also been a critical bottleneck for many other  platforms on the path to large-scale devices~\cite{burkard2023semiconductor,burkard2020superconductor,blais2021circuit,vandersypen2017interfacing}, prompting numerous  efforts  across diverse communities to overcome this challenge, such as spin shuttling~\cite{fujita2017coherent,mills2019shuttling,yoneda2021coherent,jadot2021distant,zwerver2023shuttling,noiri2022shuttling,seidler2022conveyor,langrock2023blueprint,bosco2024high} and virtual couplings enabled by  cavity photons~\cite{mi2018coherent,landig2018coherent,borjans2020resonant,yu2023strong,
harvey2022coherent,jin2012strong,bosco2022fully,
benito2019optimized,warren2019long,nigg2017superconducting,Trif2008PRB,Guido2006prb}, magnons~\cite{Daniel2013prx,flebus2019entangling,fukami2024magnon,PhysRevB.101.014416,PhysRevB.99.140403,PhysRevB.105.245310,PhysRevLett.125.247702,xue2025directional,driessen2025robust}, 
Luttinger liquids~\cite{PhysRevB.93.075301,PhysRevLett.93.126804,viennot2014stamping,PhysRevB.100.035416,PhysRevB.96.115407,PhysRevApplied.12.014030}, spin chains~\cite{Bose2003PRL,PhysRevLett.98.230503,malinowski2019fast,PhysRevLett.112.176803,baart2017coherent,PhysRevLett.126.017701} and floating gates~\cite{Daniel2012prx,PhysRevB.95.245422}.
{Among these, flying qubits offer a particularly promising strategy for scalable quantum architectures. Photons enable long-distance quantum communication in superconducting systems~\cite{ansmann2009violation}. In magnetic systems, magnons serve as natural analogs and have been proposed as carriers for quantum signal transmission via real magnon propagation~\cite{Fukami2021prx,hetenyi2022long}.
However, magnon-based schemes face fundamental limitations: their fidelity is seriously constrained by short mean free paths~\cite{Fukami2021prx}, and they do not support universal two-qubit logic, significantly restricting their usefulness in quantum computing~\cite{hetenyi2022long}.}

\begin{figure}[t!]
	\centering\includegraphics[width=0.9\linewidth]{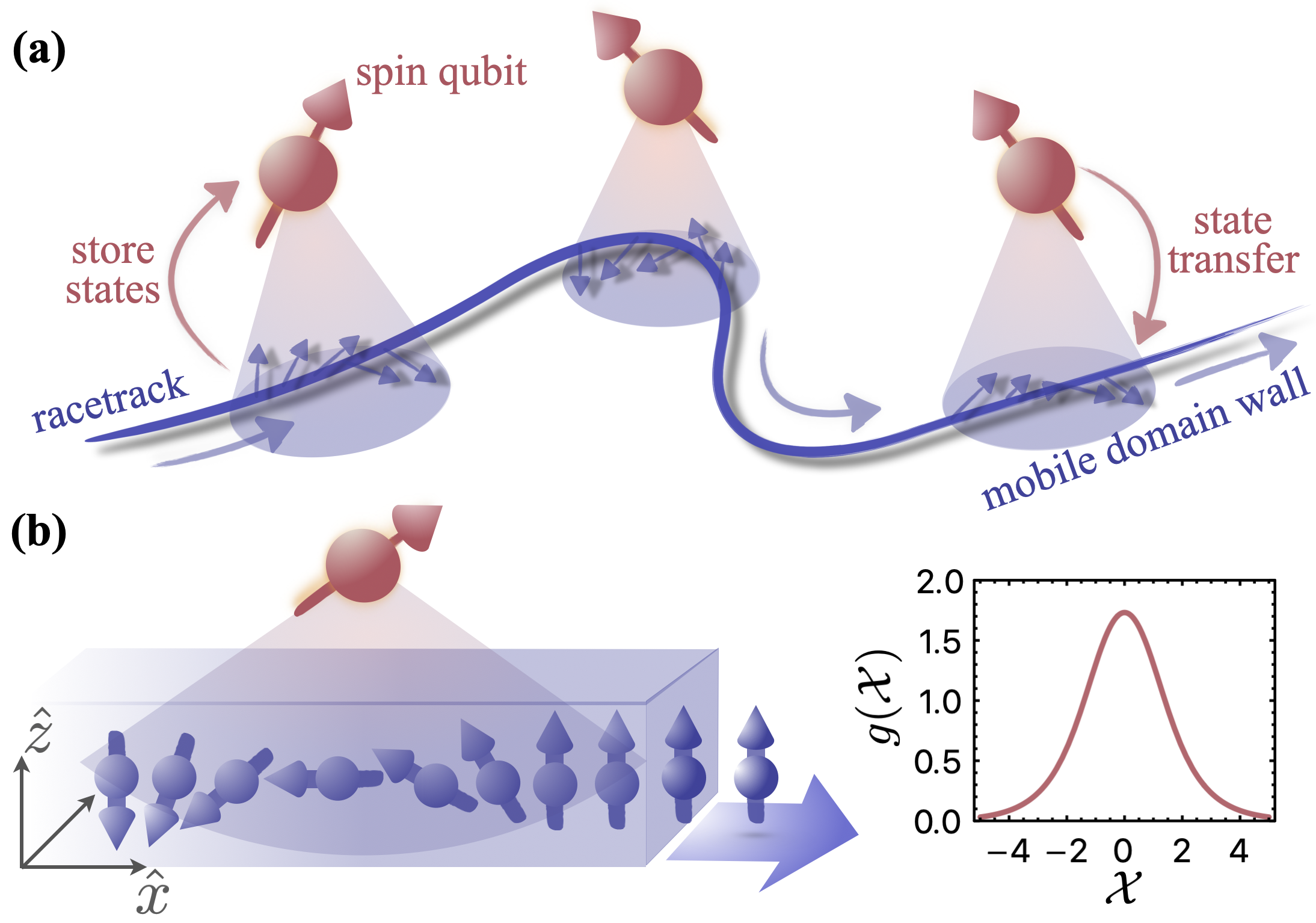}
 \caption{(a) Schematic diagram illustrating \textit{topological} quantum signal transmission in a hybrid system, where spin qubits interact with a magnetic nanowire hosting mobile DW textures.  (b) A spin qubit interacts with a DW  in the magnetic racetrack. The coupling   $g(\mathcal{X})$ varies as a function of the distance $\mathcal{X}$ between the qubit and the DW. }
  \label{fig1}
\end{figure}

{In this work, we propose a fully solid-state scheme for long-range quantum information transmission based on nanoscale mobile topological spin textures. }   We consider a hybrid system with distant spin qubits and a magnetic nanowire (racetrack), which hosts topological textures, as depicted in Fig.~\ref{fig1}(a). We illustrate the concept by examining  DWs on the track, however, this idea can also be extended to other textures such as skyrmions. 
{This  solid-state scheme offers several interesting features. First, topological textures are solitonic excitations with inherently high mobility. This has already been demonstrated experimentally~\cite{parkin2015memory}, enabling ultrafast quantum information transfer in a solid-state architecture. In contrast to bosonic excitations like magnons, which suffer from finite mean free paths and decay during propagation, topological textures preserve their nonlinear structures during motion~\cite{TopologyinMagnetism,jiprl2020,PhysRevLett.123.147203,Yqwinding,Zangprl,jivortex,Daltonenergy,quantumvortex}, allowing for high-fidelity  operations. Importantly, mobile textures can mediate long-range entangling gates between distant qubits, thus supporting universal quantum computation—overcoming a key limitation of magnon-based schemes. Furthermore, our protocol is compatible with existing racetrack memory technologies, offering a highly scalable and experimentally accessible platform. Finally, the intrinsic spin Berry phase in magnetic systems naturally induces a spin-orbit coupling, which can be harnessed to enable noise-resistant sweet spots~\cite{bosco2021hole} and implement dynamical decoupling~\cite{bosco2024high} during transport, enhancing transfer fidelity. Our work establishes a fundamentally distinct approach to quantum information transfer using topological textures, opening up new avenues for exploration.}

\textit{\textbf{Model and  racetrack-qubit interaction}}|
We begin by modeling the effective interaction between spin qubits and the low-energy manifold of the DW  in a magnetic wire.
 Let us consider  a quasi-one-dimensional  anisotropic ferrimagnetic nanowire, where each unit cell contains two spins, with microscopic Hamiltonian, $H=J\sum_{\langle i, j\rangle} \vb S_i\cdot \vb S_j- \mathcal{K}_z \sum_i(\hat{z}\cdot \vb S_i)^2 +\mathcal{K}_y\sum_i(\hat{y}\cdot \vb S_i)^2 -\hbar \sum_i\vb h\cdot \vb S_i$. Here, $J$ is the antiferromagnetic exchange coupling, whereas  $\mathcal{K}_z$ and $\mathcal{K}_y$ are both positive-valued, representing  the easy-$z$ axis and easy-$xz$ plane anisotropies, respectively. We  assume there is a  magnetic field $\vb h\equiv \gamma \vb B$   with  gyromagnetic ratio $\gamma$. This system hosts  DW textures described by the  continuous vector field $\vb n(x)$, with  $n_x(x)+in_y(x)=e^{i\phi}\sech (x-X)$ and $n_z(x)=\tanh (x-X)$~\cite{kim2023mechanics}. Here,  $\phi$ and $X$ represent the azimuthal angle and position of the DW.  All distances are expressed in units of the DW width $\lambda\equiv \sqrt{J/\mathcal{K}_z} a$ with lattice spacing $a$. 
 
 DW textures with $\phi=0\, \text{and}\, \pi$, characterized by opposite chiralities, $\mathcal{C} = (1/\pi)\int_x \hat{y} \cdot (\vb n \times \partial_x \vb n)$, are  energetically favorable with  anisotropies $\mathcal{K}_z$ and $\mathcal{K}_y$. Other energy levels are well-separated from this chirality space (the DW qubit computational space)~\cite{Zou2023prr}.  Figure~\ref{fig1}(b) depicts a profile with $\mathcal{C} = -1$ (at $\phi = \pi$).  
{We quantize both $\phi$ and $X$, then project the operators $\hat{\phi}$ and $\hat{X}$ onto the two chirality states and the orbital ground state, respectively. This yields the effective quantum Hamiltonian:
$\mathcal{H}=-t_g\tau_x/2+\varepsilon (t) \tau_z/2$
expressed in the chirality basis $\{\ket{\mathcal{C}=\pm1}\}$. Here, $\tau_{x,z}$ are Pauli matrices, $t_g$ is the chirality-state tunneling rate, and the detuning $\varepsilon(t)$ is given by  $\varepsilon=2\hbar v(t)/{\ell_{\text{so}}}-2\pi N\hbar S_eh_x$, where $S_e$ is the excess spin per unit cell of the ferrimagnetic nanowire, $N$ is the number of unit cells within the DW, and $\ell_{\text{so}}$ is the effective spin-orbit length arising from the intrinsic spin Berry phase. Crucially, the quantum treatment of $X$ renormalizes the tunneling rate $t_g$ and modifies the detuning via $v(t) = \partial_t \mathcal{X}$, where $\mathcal{X} = \langle \hat{X} \rangle$ is the expectation value of the position operator in the orbital ground state. Explicit expressions for $t_g$ and $\ell_{\text{so}}$ are given in  SM~\cite{racetrack_sm}. }

The hybrid quantum system sketched in Fig.~\ref{fig1}(a) consists of spin qubits coupled to a racetrack. We assume the spin qubits are defined within quantum dots in a 2D nonmagnetic layer deposited adjacent to the racetrack, ensuring exchange coupling~\cite{PhysRevA.57.120,hetenyi2022long}. Thus the interaction takes the form of  $V=-J_{\text{1}}\sum_{i}|\psi(\vb r_i)|^2 \vb S_i \cdot \vb*\sigma$, where $i$ runs over the region of the racetrack beneath the quantum dot, $J_1$ denotes the exchange interaction strength, $\psi(\vb r_i)$ is the orbital part of the quantum dot wavefunction, and $\vb* \sigma$ stands for the  spin qubit in the dot. To explore quantum effects, we again employ the collective coordinate quantization, projecting the dynamics onto chirality space. We   find the following effective interaction between the spin qubit and the DW low-energy dynamics~\cite{racetrack_sm}:
\( \mathcal{V}(t)=-\mathcal{J}\, g[\mathcal{X}(t)] \, \tau_z\otimes \sigma_x,  \label{eq:1}  \)
where $\mathcal{J}=J_1|\psi|^2NS_e$ is the effective coupling strength and we assumed the wavefunction  to be a constant in the dot $\psi(\vb r_i)=\psi$ for simplicity. Importantly, the time-dependent coupling envelope function is given by~\cite{racetrack_sm} 
\( g(\mathcal{X})=\sum_{k=\pm1}\arctan[\sinh(l+k\mathcal{X})],  \)
where $l$ is the radius of the dot.  Note that the interaction~\eqref{eq:1} is position-dependent, with $\mathcal{X}(t)$ representing the time-varying distance between the DW and the center of the quantum dot, while the DW moves on the track. We depict the function $g(\mathcal{X})$ in Fig.~\ref{fig1}(b) for $l=1$, which peaks when the DW is right beneath the dot and exhibits an exponential-decay tail, approaching $g(\mathcal{X})\rightarrow (4\sinh l)\exp\{-|\mathcal{X}(t)|\}$ as the DW moves away.  

We stress that the DW motion can be controlled independently of the qubit state (chirality)~\cite{racetrack_sm}. For instance, a $z$-axis magnetic field generates a force acting on the DW position $\mathcal{X}$, driving directional motion irrespective of the chirality state $\ket{\mathcal{C}=\pm 1}$. Such independent control is essential for scalable architectures.

\begin{figure}[t!]
	\centering\includegraphics[width=0.9\linewidth]{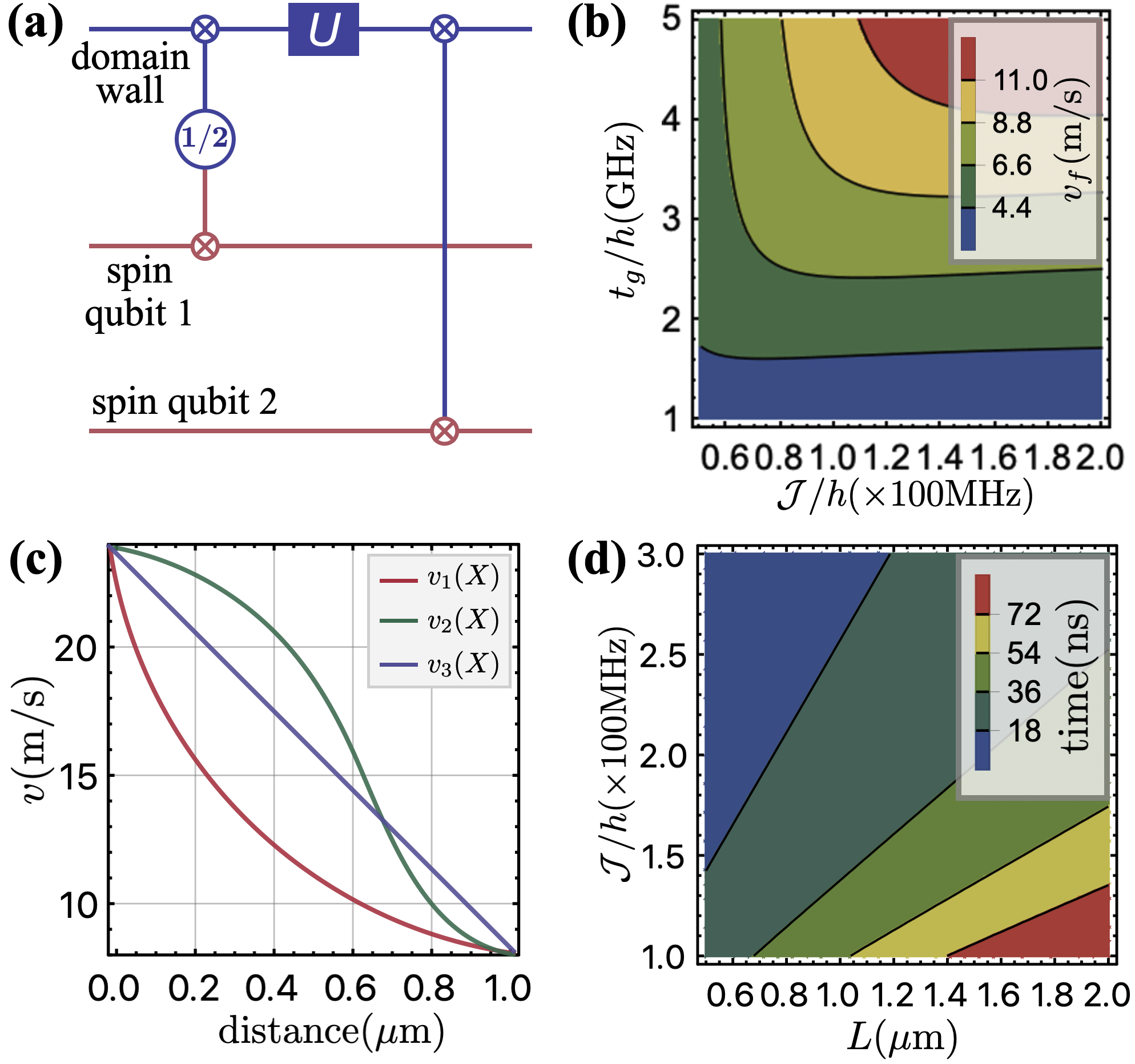}
 \caption{(a) Quantum circuit representation of the remote entanglement protocol, which  is robust to the single-qubit rotation $U$.   (b)  Optimal final DW velocity $v_f$ as a function of the qubit frequency and the racetrack-qubit coupling.  (c) Different velocity profiles on the racetrack. (d) Total operational time of the complete  protocol as a function of the coupling and separation between spin qubits. }
  \label{fig2}
\end{figure}

\textit{\textbf{Entanglement protocol}}|Here we outline  a \textit{measurement-free} remote entanglement protocol based on mobile topological textures. To this end, we examine a minimal setup involving two spin qubits $\vb*\sigma^{(i)}$ $(i=1,2)$ coupled to a  racetrack containing a  DW  described by $\vb*\tau$ in the chirality space.  The time-dependent  Hamiltonian then takes the following form,
\( \mathcal{H}(t)\!=\!\!-\!\! \sum_{i=1,2} \!\frac{\hbar \omega_s^{(i)} }{2} \sigma_y^{(i)} \!-\! \frac{t_g}{2} \tau_x\! +\! \frac{\varepsilon(t)}{2}\tau_z \!-\!\! \mathcal{J}\!\! \sum_{i=1,2}\!\! g[\mathcal{X}_i(t)] \sigma_x^{(i)}\tau_z,    \)
where $\omega_s^{(i)}$ is the spin qubit frequency determined by the magnetic field in the $y$ direction, and $\mathcal{X}_i(t)$ stands for the distance between the DW and the spin qubit $\vb*\sigma^{(i)}$. 
In the entanglement protocol, (i) starting from trivial initial  states, the DW moves beneath the first spin qubit at an optimal velocity $v_0$ facilitating a $\sqrt{\text{iSWAP}}$ gate through time-dependent interactions~\eqref{eq:1}. Then, (ii) the DW  shuttles from the first to the second  qubit, decelerating to a suitable final velocity $v_f$, during which the DW  may undergo  nontrivial unitary evolution $U$. Finally, (iii) the DW
  approaches and interacts with  the second spin qubit, realizing an iSWAP gate. These processes can be represented as the quantum circuit shown in Fig.~\ref{fig2}(a).  Importantly, this protocol operates without the need for measurements and offers  flexibility to control remote entanglement between distant  qubits on-demand.

To illustrate the key ingredients of the scheme, we first rotate the spin axis such that the relevant part of the Hamiltonian in step (i) reads  $\mathcal{H}_1(t)=-\hbar \omega^{(1)}_s\hat{\sigma}_z^{(1)}/2 -t_g\hat{\tau}_z/2+\mathcal{J} g[\mathcal{X}_1(t)]\hat\sigma_x^{(1)}\hat\tau_x$. Here, we omit the second spin qubit, which is decoupled from both the first spin qubit and the DW. We also assume operation of the DW  at the sweet spot  $\varepsilon = 0$~\cite{Zou2023prr} by applying a  constant magnetic field $h_x = v_0/(\pi NS_e\ell_{\text{so}})$. 
In the interaction picture and in the rotating wave approximation valid at resonance $t_g=\hbar\omega_s^{(1)}$, the  time-dependent interaction yields  following  evolution~\cite{racetrack_sm}:
\( \al{ \mathcal{U}\!\approx\!\cos^2\frac{\Phi}{2}+&\!\sin^2\frac{\Phi}{2}  \hat{\sigma}^{(1)}_z\hat{\tau}_z\!- \! \frac{i}{2} \sin\Phi [\hat\sigma^{(1)}_x\hat\tau_x \!+\! \hat\sigma^{(1)}_y\hat\tau_y],  \\
&\text{with} \; \;\Phi(v_0)\equiv \frac{\mathcal{J}}{\hbar}\int_0^{T} \!\!\! g[\mathcal{X}(t)]\, dt = \frac{{2\pi}\mathcal{J}}{\hbar v_0} .  } \label{eq:5}  \)
Here, we assumed that, during this step, the DW moves at a constant velocity, $\partial_t \mathcal{X}(t) = v_0$, with $T$ representing the time required for the DW to traverse the first spin qubit. Importantly, we observe that when the phase factor  $\Phi$ equals $\pi/4$, corresponding to $v_0 = {8}\mathcal{J}/\hbar $, the interaction  leads to an exact  DW-spin qubit $\sqrt{\text{iSWAP}}$ gate. With $\mathcal{J}/h\approx 100$ MHz, we have $v_0\approx 25$ m/s which is feasible in experiments~\cite{PhysRevLett.117.017202, Ryu:2013wh, Yang2015uw, Kim2017natmat,Yang:2021wy,Guan:2021vo, Blasing:2018vb,Yoshimura:2016uk}. 
 Consider the initial state of the spin qubit and the DW qubit as $\ket{\uparrow}_1$ and $\ket{\downarrow}_{\text{dw}}$, respectively. Following their interaction, they become entangled, yielding $(\ket{\uparrow}_1\ket{\downarrow}_{\text{dw}} - i \ket{\downarrow}_1\ket{\uparrow}_{\text{dw}})/\sqrt{2}$. 

To entangle the two remote qubits, we slow the DW to increase its interaction time with the second qubit, enabling a hybrid DW–spin iSWAP gate. To determine the optimal final velocity $v_{f}$, we assume a reduction by $v_r\equiv v_0-v_f$. Consequently, in step (iii), the relevant time-dependent Hamiltonian becomes $\mathcal{H}_2(t) = -\hbar \omega^{(2)}_s \hat{\sigma}_z^{(2)}/2 - t_g \hat{\tau}_z/2 + (\hbar v_r/\ell_{\text{so}}) \hat{\tau}_x + \mathcal{J} g[\mathcal{X}_2(t)] \hat{\sigma}_x^{(2)} \hat{\tau}_x$. This Hamiltonian results in a two-qubit evolution in the interaction picture similar to Eq.~\eqref{eq:5}, but importantly with a different phase factor $\tilde\Phi = {2\pi}\mathcal{J} \cos\Theta(v_r) / \hbar v_f$, with $\cos\Theta(v_r)=\{1+[2\hbar v_r/(\ell_{\text{so}} t_g)]^2\}^{-1/2}$. The desired velocity reduction $v_r$ is determined by
\(  \frac{v_0}{v_0-v_r}=\frac{2}{\cos\Theta(v_r)},    \)
where the resultant phase factor $\tilde{\Phi}$ equals $\pi/2$, corresponding to an exact iSWAP gate. We remark that the optimal final velocity $v_f$ is not the naively expected half of the initial velocity, as shown in  Fig.~\ref{fig2}(b). Instead, it must be smaller than $v_0/2$. For a qubit frequency of 3 GHz and $\mathcal{J}/h\approx  100$ MHz, the optimal final DW velocity is determined to be  $8$ m/s. 

While the DW velocity is reduced during its transition from the first to the second  qubit in step (ii), it is decoupled from both qubits, with $g(\mathcal{X}_i)=0$, and undergoes a  unitary evolution $U$ governed by $\mathcal{H}_{\text{dw}}(t)$, which depends on the changing velocity profile $v(\mathcal{X})$ shown in Fig.~\ref{fig2}(c). Different velocity profiles lead to different unitary evolutions. 
{Importantly, the protocol is immune to these variations.
As the DW approaches the second qubit, it remains maximally entangled with the first qubit. The intermediate state can be written as $\ket{\uparrow}_1\ket{\vb n}_{\text{dw}} - i\ket{\downarrow}_1\ket{-\vb n}_{\text{dw}}$, with $\bra{\vb n}\ket{-\vb n} = \bra{\downarrow}U^\dagger U\ket{\uparrow} = 0$.  Assuming the second qubit is initialized in $\ket{\uparrow}_2$, the final system state after the local iSWAP operation becomes $ \ket{\Psi_{\text{final}}}\sim  \left\{ \ket{\uparrow}_1\ket{\mathcal{S}^\dagger\vb n}_2 -i \ket{\downarrow}_1\ket{-\mathcal{S}^\dagger\vb n}_2 \right\}\otimes \ket{\uparrow}_{\text{dw}},  $ 
where $\mathcal{S}=\text{diag}(1,\, i)$ is the phase gate. After step (iii), the distant qubits form a Bell state—irrespective of the velocity profile—with the DW disentangling without projective measurements. }

{By leveraging the properties of the $\sqrt{\text{iSWAP}}$ gate, remote entanglement generation can be controllably switched on or off by simply tuning the initial state of the DW. In particular, for initial qubit states $\ket{\uparrow}_1\ket{\uparrow}_2$, flipping the DW initial state from $\ket{\downarrow}_{\text{dw}}$ to $\ket{\uparrow}_{\text{dw}}$ changes the output from a Bell state to a product state $\ket{\uparrow}_1\ket{-\mathcal{S}^\dagger\vb n}_2$, all under the same protocol. Similar entanglement switching behavior occurs for other input states, as discussed in detail in the SM~\cite{racetrack_sm}.
}

A critical parameter of the remote entanglement protocol is the operation time, which comprises three  parts: the durations of two local two-qubit gates and the transit time of the  DW. As shown in Fig.~\ref{fig2}(d), the total operational time resides in the nanosecond regime for qubit separations on the order of micrometers and coupling strengths in the MHz range. The DW-spin qubit gate time, determined by $\sim \hbar l/\mathcal{J}$, is on the order of several nanoseconds for a quantum dot size of 10 nm and $\mathcal{J}/h \approx 100$ MHz. Meanwhile, the DW transit time spans tens of nanoseconds for a spin qubit separation of approximately 1 $\mu$m. Our complete protocol, therefore, operates  within the rapid nanosecond regime.

Instead of varying the DW velocity to change its interaction time with the spin qubits, one may alternatively adjust the effective  coupling $\mathcal{J}$ using local electric fields by repositioning the spin qubit relative to the racetrack~\cite{burkard2023semiconductor}. This approach allows the realization of a $\sqrt{\text{iSWAP}}$ gate, and subsequently an iSWAP gate  of the hybrid DW-spin qubit, while maintaining a constant DW velocity throughout the entire protocol. 

{We stress that, beyond entanglement generation, mobile DWs also enable the implementation of distant two-qubit gates, such as the $\sqrt{\text{iSWAP}}$ gate~\cite{racetrack_sm}, offering a scalable route to universal quantum computing. This approach circumvents the limitations of magnon-mediated schemes~\cite{hetenyi2022long} by moving the DW back and forth: the quantum state of the first qubit is transferred to the DW, which then interacts with a distant qubit to perform an entangling gate before returning to transfer the DW state back to the first qubit. This protocol effectively realizes a $\sqrt{\text{iSWAP}}$ gate between remote spin qubits.}

\begin{figure}[t!]
	\centering\includegraphics[width=0.96\linewidth]{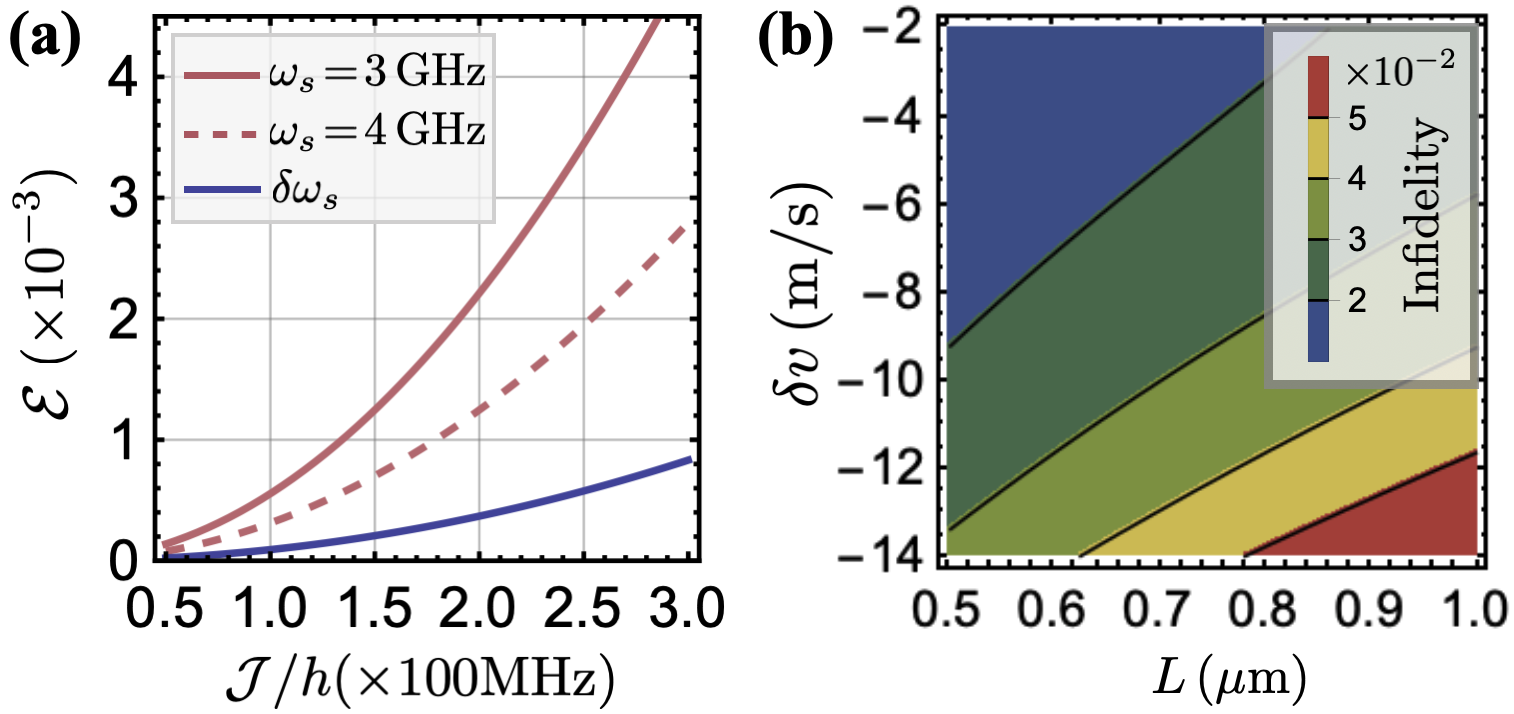}
 \caption{(a) Two-qubit gate infidelity analysis. The (dashed) red curves illustrate the error caused by spin axis tilting  for different spin qubit frequencies $\omega_s$. The purple curve depicts the error induced by qubit frequency shifts.  (b) Infidelity due to DW decoherence as a function of the velocity variation and qubit separation. }
  \label{fig3}
\end{figure}

\textit{\textbf{Analysis of  fidelity}}|To show the remote entanglement  also have high fidelity $\mathcal{F}$,  we now  focus on both coherent and incoherent errors. The coupling between the racetrack and the spin qubit leads to the desired two-qubit interaction~\eqref{eq:1} but also induces a time-dependent magnetic field in the $z$ direction on the spin qubit, $h_z[\mathcal{X}(t)] = \mathcal{J}\ln [\cosh(l-\mathcal{X})/ \cosh(l+\mathcal{X})]$. This field causes the spin qubit axis to  tilt by a small angle $\delta\theta \approx 2h_z/\hbar\omega_s$ and induces a small  shift in the qubit frequency, $\delta\omega_s = \sqrt{\omega_s^2 + (2h_z/\hbar)^2} - \omega_s$.  

Importantly,  we find that the resultant infidelity of the hybrid DW-spin qubit gate, $\mathcal{E} = 1 - \mathcal{F}$, is always bounded from above and is analytically given by~\cite{racetrack_sm}
\( \mathcal{E} \! \leq \! 2\sin^2\frac{\delta\theta}{2} \!+\! \int^T_0\!\! \frac{dt}{2\hbar} \int^T_0\!\!  \frac{ds}{2\hbar} \check{\mathcal{J}}(t)\check{\mathcal{J}}(s) \cos[{2}(\Phi(t)-\Phi(s))], \label{eq:6} \)
where $T$  is again the  gate time, $\Phi(t)=\mathcal{J}\int^t_0g[\mathcal{X}(\tau)]d\tau/\hbar$, and $\check{\mathcal{J}}(t)=\mathcal{J}g[\mathcal{X}(t)]\sin(\delta\omega_s t)$. The first term of Eq.~\eqref{eq:6} represents the maximal infidelity due to the tilting of the  qubit axis. The (dashed) red curves in Fig.~\ref{fig3}(a) show the corresponding  value for different spin qubit frequencies $\omega_s$, as a function of the coupling $\mathcal{J}$, which is on the order of $10^{-3}$. The purple curve in Fig.~\ref{fig3}(a) illustrates the infidelity induced by the qubit frequency shift, corresponding to the second term in Eq.~\eqref{eq:6}, which is estimated to be on the order of $10^{-4}\sim 10^{-3}$ for the racetrack-qubit coupling in the MHz regime. 

We now focus on the infidelity induced by DW decoherence. With the DW velocity profile $v(\mathcal{X}) = v_0 + \delta v[\mathcal{X}(t)]$ in general on the racetrack, we show that the resultant infidelity is~\cite{racetrack_sm}:
\(\mathcal{E}\!\approx\! \frac{1}{8\hbar^2} \!\! \int^\infty_{-\infty}\!\! \frac{d\omega}{2\pi} S(\omega) |\vb*{\mathcal{M}} (\omega)|^2,\, \text{with}\; \vb*{\mathcal{M}}\!=\!\!\!\int^{\bar{T}}_0 \!\!\! d\tau\, e^{-i\omega \tau} \!\vb m(\tau),  \)
where $S(\omega)/\hbar^2\!\!=\!\!\alpha N\omega \coth[\hbar\omega/(2k_BT_0)]$ is the noise spectral function with Gilbert damping  $\alpha$ and temperature $T_0$, $\bar{T}$ is the total operational time, and   $|\vb*{\mathcal{M}}(\omega)|^2$ is the effective filter function with  $\vb m(t)\!=\!(\sin \theta(t)\cos\tilde{\Phi}(t), \sin \theta(t)\sin\tilde{\Phi}(t), \cos \theta(t))$. Here, $\theta(t)$ is determined by $\tan\theta(t)=-t_g \ell_{\text{so}}/[2\hbar\delta v(t)]$ and $\tilde{\Phi}(t)=\int^t_0d\tau \,t_g/[\hbar \sin\theta(\tau)]$. The detailed derivations are presented in SM~\cite{racetrack_sm}. Assuming a DW coherence time ($T_2$)  of 0.5 $\mu$s and a constant $\delta v$ for simplicity~\cite{psaroudaki2023skyrmion}, we show the infidelity as a function of $\delta v$ and the separation $L$ between  qubits in Fig.~\ref{fig3}(b).
The error is generally on the order of $10^{-2}$  for a velocity variation  $\delta v \sim -10$ m/s and a qubit separation of micrometers. {The infidelity from spin qubit decoherence can be estimated as $\mathcal{E} \approx \bar{T}/T_2 \approx 5\times 10^{-3}$, with $\bar{T}\approx 50\,\text{ns}$ [see Fig.~\ref{fig2}(d)] and spin qubit lifetime $T_2\approx 10\,\mu\text{s}$~\cite{wu2025simultaneous,burkard2023semiconductor}, which is about an order of magnitude smaller than the infidelity from DW decoherence.}

{We contrast our proposed scheme with magnon-based approaches in solid-state systems, which generate remote entanglement by transmitting real magnons—akin to the role of photons in optical systems. However, the conservation of  magnons  relies on an underlying U(1) symmetry that is easily broken by imperfections. This results in magnon decay during propagation and severely limits entanglement fidelity. In stark contrast, topological spin textures  are protected by topology and remain stable during motion, making high-fidelity entanglement generation  over long distances possible.}

\textit{\textbf{Quantum station for racetrack}}|In the hybrid quantum system depicted in Fig.~\ref{fig1}(a), spin qubits also nicely serve as quantum stations for the racetrack, not only storing but also facilitating the exchange of DW quantum states, which is crucial for developing a scalable quantum communication system. We consider a single spin qubit coupled to a racetrack containing multiple DWs. Assuming the quantum station—the spin qubit—is prepared in state $\ket{\uparrow}_{s}$, and DW $A$ is in a general unknown mixed  state $\rho_A = \sum_i p_i\ket{\psi_i}\bra{\psi_i}$, we propose moving the DW at an optimal velocity $\tilde{v} = 4\mathcal{J}/ \hbar$ to store $\rho_A$ in the spin qubit. This motion facilitates a unitary evolution that precisely executes an iSWAP gate acting on the hybrid DW-spin qubit, leading to the final state $\rho_s = \mathcal{S}^\dagger \rho_A \mathcal{S}$ of the spin qubit. By subsequently applying a phase gate, we effectively store the exact unknown state $\rho_A$ into the spin qubit as shown in Fig.~\ref{fig1}(a). 

When another DW $B$ on the racetrack in state $\ket{\uparrow}_B$ passes by the  station at the velocity $\tilde{v}$ at later time, the time-dependent interaction with the spin qubit  again realizes another  iSWAP gate, leading to the final state of DW $B$ as $\rho_B = \mathcal{S}^\dagger \rho_A \mathcal{S}$. Therefore,  the spin qubit facilitates the transfer of quantum states between fast-moving DWs on the same racetrack, enabling quantum information exchange among different topological spin textures on the same racetrack, which would otherwise be a challenging task. In the dual picture, DWs also act as \textit{mobile} quantum stations for spin qubits, facilitating long-distance spin-qubit state transfer. We note that by moving the first DW at a slower velocity or adjusting the  coupling $\mathcal{J}$ with local electric fields to enable a hybrid DW-spin qubit $\sqrt{\text{iSWAP}}$ gate, we can also entangle different DWs on the same track.

{\textit{\textbf{Discussion}}|A key requirement for realizing DW–based quantum information transport is maintaining nanoscale DW widths for reliable qubit encoding. Atomically thin magnetic multilayers such as Co/Ni and CoFeB are promising candidates due to their strong magnetic anisotropy, supporting sub-10 nm DWs and demonstrated velocities exceeding 100 m/s~\cite{parkin2015memory}. Two-dimensional van der Waals magnets, such as $\text{Fe}_3\text{GeTe}_2$, also offer attractive properties, including low damping for improved coherence and reduced current requirements for DW motion, minimizing heating~\cite{zhang2024current}. 

In parallel, spin qubits confined in silicon or germanium quantum dots provide a compatible and scalable platform for integration with  racetracks~\cite{PhysRevA.57.120}. Micromagnets are already widely used in spin qubit manipulation~\cite{burkard2023semiconductor}, and in the present context, such qubits can couple to DW via  dipole or exchange interactions. Together, these material platforms offer a feasible and promising route for realizing the proposed architecture.}

\begin{acknowledgments}
\textit{Acknowledgments}|We thank Banabir Pal and Stuart Parkin  for insightful discussions.
This work was supported by the Georg H. Endress Foundation and by the Swiss National Science Foundation, NCCR SPIN (Grant Number: 51NF40-180604). 
\end{acknowledgments}


%

\end{document}


\title{ Supplemental Material for \\``Topological Spin Textures Enabling Quantum Transmission"}

\author{Ji Zou}
\affiliation{Department of Physics, University of Basel, Klingelbergstrasse 82, 4056 Basel, Switzerland}
\author{Stefano Bosco}
\affiliation{QuTech, Delft University of Technology, Lorentzweg 1, 2628 CJ Delft, The Netherlands}
\author{Jelena Klinovaja}
\author{Daniel Loss}
\affiliation{Department of Physics, University of Basel, Klingelbergstrasse 82, 4056 Basel, Switzerland}

\begin{abstract}
\end{abstract}
\maketitle

\tableofcontents{}

\subsection{(I) Deriving the effective coupling between spin qubits and domain walls}
In this section, we present the derivation of the interaction between spin qubits and domain walls, the Equations (1) and (2) in the main text. We begin with a brief review of the derivation of the domain-wall low-energy dynamics (the construction of the domain-wall chirality qubit Hamiltonian). The microscopic Hamiltonian of the anisotropic ferrimagnetic nanowire is $H=J\sum_{\langle i, j\rangle} \vb S_i\cdot \vb S_j- \mathcal{K}_z \sum_i(\hat{z}\cdot \vb S_i)^2 +\mathcal{K}_y\sum_i(\hat{y}\cdot \vb S_i)^2 -\hbar \sum_i\vb h\cdot \vb S_i$, as we presented in the main text.  In the construction, we only need to apply a magnetic field in the  $y$  direction, so we assume $\vb{h} = h_y\hat{y}$. The quantum dynamics of the system can be described as a  continuous vector field $\vb n(x)$ in the spin path integral formalism. Since our main focus is the domain wall texture, we assume the boundary condition $n_z(\pm \infty)=\pm \eta$ where $\eta=\pm 1$. The domain wall configuration is given by~\cite{kim2023mechanics}
\( \al{  n_x(x) &=\cos\phi \sech (x-X), \\
         n_y(x)&=\sin\phi \sech(x-X), \\
        n_z(x)&=\eta \tanh(x-X),   }  \label{S1} \)
Here, $X$ and  $\phi$ represent the spatial position and the azimuthal angle in spin space of the DW.   All distances are expressed in terms of the DW width $\lambda\equiv \sqrt{J/\mathcal{K}_z} a$ with lattice spacing $a$. 
 
We are particularly interested in the low-energy dynamics of the DW captured by the two soft modes $\vb*{\psi}\equiv ({X}, \phi)$. Within the collective coordinate approach, the Lagrangian governing such dynamics is given by
\(  \mathcal{L}(\vb*\psi)= \frac{\mathcal{M} }{2} (\partial_t \vb*\psi)^2 -\vb*{\mathcal{A}} \cdot \partial_t \vb*\psi- \mathcal{F}(\vb*\psi), \)
where the effective mass is $\mathcal{M}=N\hbar^2/2J$ with $N$ being the number of unit cells within the DW. The effective gauge field reads $\vb*{\mathcal{A}}=(\pi \mathcal{M}h_y\cos\phi, 4N\hbar S_e X)/2$, where $S_e$ represents the excess spin within the unit cell. We point out that this emergent gauge field arises from the spin Berry phase, accumulated by the net magnetization induced by the applied magnetic field $h_y$ and by the excess spin. The potential energy is $\mathcal{F}=2N\mathcal{K}_y(\sin^2\phi -2b_y\sin\phi)+\mathcal{M}\omega_0^2 X^2/2$. For clarity, $b_y$ is defined as the dimensionless magnetic field $\pi\hbar S_eh_y/4\mathcal{K}_y$, and we introduced a confining potential characterized by a harmonic length $\ell_0 = \sqrt{\hbar/(\mathcal{M}\omega_0)}$. Importantly, the double-well potential for $\phi$ favoring domain wall textures close to  $\phi=0$ or $ \pi$, characterized by opposite chiralities. Note that due to the finite value of $b_y$, the favorable profiles are slightly shifted from $\phi=0, \pi$. Both chirality states possess equal energy and are separated by a tunnel barrier, $\mathcal{V}_0 = 2N\mathcal{K}_y(1-b_y)^2$~\cite{Zou2023prr}. 

{We quantize the two degrees of freedom, $\hat{\phi}$ and $\hat{X}$. The $\hat{\phi}$ dynamics is projected onto the two chirality states, while the spatial dynamics governed by $\hat{X}$ is projected onto the orbital ground state.} This yields the effective description for the domain wall: $\mathcal{H}=-t_g\tau_x/2$ in the chirality basis $\ket{\mathcal{C}=\pm1}$ where $\tau_x$ is the Pauli matrix and the tunneling splitting reads~\cite{Zou2023prr}:
\( t_g \approx 4\hbar \omega_s \sqrt{ \frac{2\mathcal{V}_0}{\pi \hbar \omega_s} } \exp{-\left(\frac{4\mathcal{V}_0}{\hbar \omega_s} +\frac{\ell_0}{\ell_{\text{so}}} \right) }.  \)
Here, $\hbar \omega_s = 2\sqrt{2J\mathcal{K}_y(1 - b_y^2)}$ quantifies the energy gap from the low-energy subspace to the higher energy sector. We define the effective spin-orbital length $\ell_{\text{so}} \approx (2\hbar/\pi)/(\mathcal{M}h_y + 2N\hbar S_e)$, which represents the distance the domain wall needs to travel to flip its chirality, an effect driven by the effective gauge field $\vb*{\mathcal{A}}$. {We stress that the spatial degree of freedom is treated quantum mechanically, and it renormalizes the tunneling rate by a factor of $ \exp(-\ell_0 /\ell_{\text{so}}) $, as shown above. In the presence of a magnetic field along the racetrack, or when the domain wall is in motion, the two chirality states are no longer energetically equivalent. This asymmetry introduces a detuning term $\varepsilon \tau_z/2$ into the Hamiltonian, where $\varepsilon = 2\hbar v(t)/\ell_{\text{so}} - 8N\mathcal{K}_y b_x$. Here, $v \equiv \partial_t \langle \hat{X} \rangle$ denotes the velocity of the domain wall, with $\langle \hat{X} \rangle$ representing the expectation value of the position operator in the orbital ground state. } The projection of the $\hat\phi$ dynamics onto the chirality space can be summarized as $\cos\phi \rightarrow \gamma_z \tau_z$ and $\sin\phi\rightarrow \gamma_0I+\gamma_x\tau_x$. We work in the same parameter regime as Ref.~\cite{Zou2023prr}, and we have $\gamma_z=0.9, \gamma_0=0.14, \gamma_x=2.7\times 10^{-3}$.

To transfer information between the spin qubit and the domain wall qubit, we require a sizable coupling between the spin qubit and the racetrack. To achieve this, we assume that the spin qubits are defined by electrostatic gates in a 2D nonmagnetic layer deposited adjacent to  the magnetic racetrack, allowing the spin qubits to be exchange-coupled to the racetrack. We also assume the quantum dot size $l$ to be around 10 nm in both the $x$ and $y$ directions, which is feasible in experiments. The coupling between the spin qubit $\vb* \sigma$ and the spins within the magnetic nanowire is
\( \al{ V&=-J_1\sum_{i\in\text{dot}} |\psi(\vb r_i)|^2 \vb S_i\cdot \vb* \sigma \approx -J_1 |\psi|^2 \sum_{i\in\text{dot}}  \vb S_i\cdot \vb* \sigma = -J_1 |\psi|^2 NS_e \int^l_{-l} dx\, \vb n(x) \cdot \vb* \sigma  \\
           &= - J_1 |\psi|^2 NS_e \left[  \left(  \int^l_{-l} dx \, n_x (x) \right) \sigma_x  +  \left(  \int^l_{-l}  dx \, n_y (x) \right) \sigma_y + \left(  \int^l_{-l}  dx \, n_z(x)  \right) \sigma_z   \right].     }  \)
      Here $J_1$ is the exchange coupling strength which can be quite large; the summation $\sum_i$ only runs over the regime in the racetrack covered by the quantum dot; $|\psi(\mathbf{r}_i)|^2$ is the orbital part of the quantum dot wave function, which we approximate with a constant $|\psi|^2$ in the dot. We have $\sum_i |\psi|^2 = 1$, indicating that $|\psi|^2$ should be inversely proportional to the quantum dot size. We have again rescaled our coordinate such that we measure the distance in the unit of domain wall width $\lambda\approx 5\, \text{nm}$ for convenience. For a quantum dot size of 10 nm,  we have $l=1$ (with $l$ being radius of the dot). $N$ and $S_e$ are the total number of unit cells within the domain wall and the net spin within one unit cell, which we assume to be $N S_e \sim 1$. Now we perform the integral and then project $\phi$ onto the chirality basis. For the $x$ component, we have
      \(  \left(  \int^l_{-l} dx \, n_x (x) \right) \sigma_x =  \sigma_x \cos\phi \int^l_{-l} dx\, \sech(x-\mathcal{X}) = g(\mathcal{X})\sigma_x \cos\phi  \rightarrow g(\mathcal{X}) \gamma_z\, \tau_z\otimes \sigma_x. \)
      Here $g(\mathcal{X})\!=\!\arctan[\sinh(l+\mathcal{X})]+\arctan[\sinh(l-\mathcal{X})]$. Similarly for the $y$ and $z$  components, we obtain $\mathcal{J}g(\mathcal{X})(\gamma_0+\gamma_x \tau_x)\otimes \sigma_y$ and $ h_z(\mathcal{X})  \sigma_z$, where the effective field due to the racetrack acting on the spin qubit is: 
      \( h_z(\mathcal{X})=\eta \mathcal{J} \ln \left[ \frac{\cosh(l-\mathcal{X})}{\cosh(l+\mathcal{X})}\right]. \)
      Here we introduced $\mathcal{J}=J_1|\psi|^2NS_e$ and $\eta=\pm 1$, again, stands for the boundary condition of the domain wall texture that introduced in Eq.~\eqref{S1}.
      We note that $h_z(\mathcal{X}=0)=0$, indicating that the effect of the racetrack ($z$ component) averages to zero when the domain wall is right beneath the quantum dot. Then the effective interaction between the spin qubit and the domain wall qubit is given by (involving both $\vb*\tau$ and $\vb*\sigma$)
      \(\mathcal{V}(\mathcal{X}) = -\mathcal{J} g(\mathcal{X}) \left( \gamma_z \tau_z \otimes \sigma_x +\gamma_x \tau_x\otimes \sigma_y  \right) \approx  -\mathcal{J} g(\mathcal{X}) \tau_z \otimes \sigma_x, \)
which is the Equation (1) in the main text. Here we have approximated $\gamma_z=0.9\approx 1$ and $\gamma_x\sim 10^{-3}\approx 0$. We remark that we have dropped single qubit terms, such as the term $h_z(\mathcal{X})\sigma_z$, which will be taken into account later in the analysis of the protocol fidelity.

\medskip
{ \subsection{(II) Controlling domain wall motion independently of qubit state}
It is important to note that the domain wall motion (both its direction and speed) can be controlled independently of the qubit state. Several methods can achieve this decoupled control; for illustration, we consider the simplest case: applying a magnetic field $H_z$ along the $z$-direction. This magnetic field influences the DW dynamics in two distinct ways. First, it induces a net magnetization in the ferrimagnet, contributing a spin Berry phase that manifests as a shift in the gauge potential for $\phi$; second, it couples to the net magnetization via the Zeeman interaction, introducing a linear potential in $X$ within the collective-coordinate framework:
\( \mathcal{A}_\phi\rightarrow \tilde{\mathcal{A}_\phi }=  \mathcal{A}_\phi + MH_z,\;\;\;\text{and} \;\;\; \mathcal{F}(\vb*\psi)\rightarrow  \mathcal{F}(\vb*\psi) -2\hbar S_eNH_zX. \)
When we quantize both $\phi$ and $X$ and express the system in Hamiltonian formalism, two terms involving the magnetic field $H_z$ are given by:
\(   \hat{H}= \frac{(\hat{P}_\phi+\mathcal{A}_\phi + MH_z )^2}{2M} -2\hbar S_eNH_z\hat{X}. \)
It should be clear that the $H_z$ in the first term can be gauged away by a gauge transformation $\hat U = \exp(-i \hat{\phi} M H_z)$, which effectively shifts $\hat{P}_\phi \rightarrow \hat{P}_\phi - M H_z$. This is possible because $\phi$ is well localized within the double-well potential in our case, making it effectively non-compact. Then, the effect of $H_z$ lies solely in the last term in the expression above, which generates a force, $-\langle \partial \hat{V}/\partial \hat{X} \rangle = 2\hbar S_e N H_z$ according to Ehrenfest theorem, acting on the domain wall position $\mathcal{X} = \langle \hat{X} \rangle$, independent of the chirality of the state.
}

\medskip
\subsection{(III) Effective unitary evolution under the racetrack-qubit interaction}
In this section, we present the effect of the position-dependent coupling  when we move the domain wall across the spin qubit and derive the Eqs. (4) and (5) in the main text. To this end, we consider a single spin qubit coupled to a racetrack with following effective Hamiltonian:
\( \mathcal{H}(t)\approx - \frac{\hbar \omega_s}{2} \sigma_y - \frac{t_g}{2} \tau_x + \frac{\varepsilon(t)}{2}\tau_z - \mathcal{J} g[\mathcal{X}(t)] \sigma_x \otimes \tau_z.    \)
Here, the first term is due to the magnetic field we apply in the $y$ direction. We have dropped the term $\mathcal{J}h_z(\mathcal{X}) \sigma_z$ for now, but we will return to it to evaluate the induced coherent error. We note that, though the Hamiltonian above is a $4\times 4$ matrix, the dynamics is not exactly solvable due to the nontrivial time-dependent interaction.  We will assume the domain wall moves at a constant velocity $\partial_t \mathcal{X}(t)=v$ during its interaction with the spin qubit.

\textbf{Case I: Dynamics with vanishing domain-wall qubit detuning.}| We now work at the sweet spot $\varepsilon=0$, which can be achieved by turning on a small magnetic field in the $x$ direction. When the domain wall velocity is $v=10\, \text{m/s}$, the magnetic field is about $B_x\approx 8\, \text{mT}$.
We  first rotate the spin axis: $\tau_x\rightarrow \hat{\tau}_z, \tau_z\rightarrow -\hat{\tau}_x$ and $\sigma_y\rightarrow \hat{\sigma}_z, \sigma_z\rightarrow -\hat{\sigma}_y$. Then the Hamiltonian reads 
\(  \mathcal{H}(t)=- \frac{\hbar \omega_s}{2} \hat{\sigma}_z - \frac{t_g}{2} \hat{\tau}_z +\mathcal{J} g[\mathcal{X}(t)] \hat{\sigma}_x \otimes \hat{\tau}_x.  \)
We perform the unitary transformation: 
\(  U_1(t)= \exp{\frac{i\omega_s t  }{2}\hat{\sigma}_z}\otimes \exp{\frac{it_g t  }{2}\hat{\tau}_z}, \; \;\text{and}\;\; \mathcal{H}\rightarrow \tilde{\mathcal{H}}=U_1^\dagger\mathcal{H} U_1 +i\hbar \partial_t U_1^\dagger U_1.   \)
In the following, assuming $t_g=\hbar \omega_s$, we have
\(\al{ \tilde{\mathcal{H}}(t) &= \mathcal{J} g[\mathcal{X}(t)]  U_1^\dagger \hat{\sigma}_x \otimes \hat{\tau}_x U_1 = \mathcal{J} g[\mathcal{X}(t)] \left\{  \hat{\sigma}_+\hat{\tau}_+ \exp[-i(\omega_s+t_g/\hbar)t] + \hat{\sigma}_+ \hat{\tau}_- \exp[i(t_g/\hbar -\omega_s)t]  +\text{h.c.}  \right\}   \\
 &\approx \mathcal{J} g[\mathcal{X}(t)] ( \hat{\sigma}_+\otimes \hat{\tau}_- +\hat{\sigma}_- \otimes \hat{\tau}_+  ) = \frac{\mathcal{J} g[\mathcal{X}(t)] }{2}  (\hat{\sigma}_x\otimes \hat{\tau}_x +\hat{\sigma}_y \otimes \hat{\tau}_y ),   } \)
where we have used $t_g=\hbar \omega_s$ and the rotating wave approximation from the first to the second line (which is valid since the coupling strength $\mathcal{J}$ is in the MHz regime, much smaller than the qubit frequency). Then the evolution due to the Hamiltonian is given by 
\( \al{ \mathcal{U}&=\exp{-\frac{i}{\hbar} \int^T_0 d\tau\, \tilde{\mathcal{H}}(\tau)   } =\cos^2\frac{\Phi}{2}+\!\sin^2\frac{\Phi}{2}  \hat{\sigma}_z \otimes\hat{\tau}_z\!- \! \frac{i}{2} \sin\Phi [\hat\sigma_x \otimes \hat\tau_x \!+\! \hat\sigma_y \otimes\hat\tau_y]  \\
&= \mqty[ 1 & 0 & 0 & 0 \\ 0 & \cos\Phi  & -i\sin\Phi & 0 \\ 0 & -i\sin\Phi & \cos \Phi & 0 \\ 0& 0& 0& 1   ], \;\;\text{where}\;\; \Phi= \frac{\mathcal{J}}{\hbar}\int^T_0 d\tau g[\mathcal{X}(\tau)] = \frac{\mathcal{J}}{\hbar v} \int g(\mathcal{X}) d\mathcal{X} = \frac{ {2\pi} \mathcal{J}}{\hbar v},  } \)
where the basis of the matrix is $\ket{\uparrow\uparrow}, \ket{\uparrow\downarrow}, \ket{\downarrow\uparrow}, \ket{\downarrow\downarrow}$. Here, we have used the fact that the domain wall moves at a constant velocity $v$ while it can also be generalized to a time-dependent velocity profile, as long as the motion remains adiabatic with respect to the excited DW states above the two chirality states, $v\leq 100\,\text{m/s}$~\cite{Zou2023prr}.  Here, $T$ is the time for the domain wall to traverse the spin qubit. This is the Equation (4) in the main text. 

\textbf{Case II: Dynamics with finite domain-wall qubit detuning.}|With a fixed global magnetic field along the racetrack  $B_x$  in the entanglement protocol, a domain-wall qubit detuning  $\varepsilon \neq 0$  arises when the domain wall moves at a different velocity. We assume the velocity is changed from $v_0$ to $v_f$. In this case, the Hamiltonian takes the form of 
\(   \mathcal{H}(t)=- \frac{\hbar \omega_s}{2} \hat{\sigma}_z - \frac{t_g}{2} \hat{\tau}_z + \frac{\hbar v_r}{\ell_{\text{so}}} \hat{\tau}_x  +\mathcal{J} g[\mathcal{X}(t)] \hat{\sigma}_x \otimes \hat{\tau}_x,  \)
where $v_r=v_0-v_f$. We first rotate the spin space of $\vb* \tau$ in $xz$-plane with the rotation matrix: 
\( \mathcal{R}(\Theta)=\mqty[  \cos \Theta  &  -\sin\Theta  \\ \sin \Theta & \cos \Theta   ], \;\; \text{with}\;\; \tan\Theta=\frac{2\hbar v_r}{t_g \ell_{\text{so}}}.   \)
We then have the following Hamiltonian: 
\(  \mathcal{H}(t)= - \frac{\hbar \omega_s}{2} \hat{\sigma}_z - \frac{\Delta}{2} \hat{\tau}_z + \mathcal{J} g[\mathcal{X}(t)] \hat{\sigma}_x \otimes \left( \cos\Theta \hat{\tau}_x- \sin\Theta \hat{\tau}_z  \right),  \)
where $\Delta=\sqrt{t_g^2+(2\hbar v_r/\ell_{\text{so}})^2}$. We again tune the spin qubit frequency such that they are on resonance $\hbar \omega_s= \Delta$. Under the rotating wave approximation in the interaction picture, we have the following Hamiltonian: 
\(  \tilde{\mathcal{H}}(t)=  \frac{\mathcal{J} \cos\Theta g[\mathcal{X}(t)] }{2}  (\hat{\sigma}_x\otimes \hat{\tau}_x +\hat{\sigma}_y \otimes \hat{\tau}_y ).   \)
We can see two differences compared to Case I: the domain-wall velocity is now  $v_f = v_0 - v_r$, and the effective coupling changes to $\mathcal{J} \rightarrow \mathcal{J} \cos \Theta$. 

In the main text, we want the domain wall to initially move at velocity  $v_0$  to achieve a $\sqrt{\text{iSWAP}}$ gate ($\Phi = \pi/4$), and later at velocity $v_f$  to realize an iSWAP gate ($\Phi = \pi/2$), which would require: 
\(  \frac{2\pi\mathcal{J}}{\hbar v_0}= \frac{\pi}{4} \;\; \;\;\text{and} \;\; \;\; \frac{2\pi\mathcal{J} \cos\Theta}{\hbar(v_0-v_r)}= \frac{\pi}{2},   \)
where we note the angle $\Theta$ also depends on $v_r$. From these two equations, we deduce the Equation (5) in the main text. 

\subsection*{(IV) Discussion of different initial states and the robustness of the entanglement protocol}
In this section, we discuss in detail the robustness of the entanglement protocol and present the results for different initial states. 

\textbf{Case I.}|In the main text, we consider the initial states to be $\ket{\uparrow}_1\otimes\ket{\uparrow}_2\otimes \ket{\downarrow}_{\text{dw}}$. The entangling gate between the first spin qubit and the domain wall yields the state:
\(  \ket{\Psi_1} = \frac{ \ket{\uparrow}_1\ket{\downarrow}_{\text{dw}}-i\ket{\downarrow}_1\ket{\uparrow}_{\text{dw}}    }{2}\otimes \ket{\uparrow}_2.   \)
When the domain wall is moving from the first spin qubit to the second, its velocity is reduced from $v_0$ to $v_f$ during which its dynamics is governed by a time-dependent Hamiltonian, leading to a unitary evolution: 
\( U=\mathcal{T} \exp{ \int d\tau \mathcal{H}_{\text{dw}}(\tau)  }, \;\; \text{with}\;\; \mathcal{H}_{\text{dw}}(\tau)= -\frac{t_g}{2}\hat{\tau}_z + \frac{\hbar [v_0-v(\tau)]}{\ell_{\text{so}}} \hat{\tau}_x.   \)
Here, $\mathcal{T}$ is the time-ordering operator and $v(\tau)$ is the domain wall velocity at time $\tau$. It is clear that the exact evolution depends on how we reduce the velocity $v(\tau)$ from $v_0$ to $v_f$. Importantly, different evolutions do not change the entanglement between the first spin qubit and the domain wall: 
\(  \ket{\Psi_1} \rightarrow \ket{\Psi_2} = \frac{ \ket{\uparrow}_1\ket{\vb n}_{\text{dw}}-i\ket{\downarrow}_1\ket{-\vb n}_{\text{dw}}    }{2}\otimes \ket{\uparrow}_2, \;\; \text{with}\;\; \ket{\vb n}_{\text{dw}} =U\ket{\downarrow}, \;\text{and}\;   \ket{-\vb n}_{\text{dw}} =U\ket{\uparrow}. \)
This robustness is rooted in the fact that a local unitary transformation does not change the entanglement between two objects. To see the effect of the iSWAP gate, we assume $\ket{\vb n}_{\text{dw}} = a \ket{\uparrow}_{\text{dw}}  + b \ket{\downarrow}_{\text{dw}} $ and $\ket{-\vb n}_{\text{dw}}= b^* \ket{\uparrow}_{\text{dw}} - a^* \ket{\downarrow}_{\text{dw}}$. Recall that we have $\ket{\uparrow\downarrow}\rightarrow -i \ket{\downarrow \uparrow}$ and $\ket{\downarrow\uparrow}\rightarrow -i\ket{\uparrow\downarrow}$ under the iSWAP gate. Then it is clear that, after the iSWAP gate, we have
\(  \ket{\vb n}_{\text{dw}}\otimes \ket{\uparrow}_2\rightarrow \ket{\uparrow}_{\text{dw}}\otimes \ket{\mathcal{S}^\dagger\vb n}_2, \;\;\text{and}\;\;  \ket{-\vb n}_{\text{dw}}\otimes \ket{\uparrow}_2\rightarrow   \ket{\uparrow}_{\text{dw}}\otimes \ket{-\mathcal{S}^\dagger\vb n}_2, \;\;\text{with phase gate}\;\;\mathcal{S}=\mqty[1 & 0 \\ 0 & i]. \)
Therefore, with the initial state $\ket{\uparrow}_1\otimes\ket{\uparrow}_2\otimes \ket{\downarrow}_{\text{dw}}$, the final state is $\left\{ \ket{\uparrow}_1\ket{\mathcal{S}^\dagger\vb n}_2 -i \ket{\downarrow}_1\ket{-\mathcal{S}^\dagger\vb n}_2 \right\}\otimes \ket{\uparrow}_{\text{dw}}$. When we flip the initial state of the domain wall from $\ket{\downarrow}_{\text{dw}}$ to $\ket{\uparrow}_{\text{dw}}$, the first spin qubit will not be entangled with the domain wall under the same protocol. When the domain wall is approaching the second spin qubit, the system state is $\ket{\uparrow}_1\otimes \ket{-\vb n}_{\text{dw}}\otimes \ket{\uparrow}_2$. Thus, after the iSWAP gate, the final state is $\ket{\uparrow}_1\otimes\ket{-\mathcal{S}^\dagger\vb n}_2\otimes \ket{\uparrow}_{\text{dw}}$, which is a trivial product state. We can summarize as 
\(  \ket{\uparrow}_1\otimes\ket{\uparrow}_2\otimes \ket{\downarrow}_{\text{dw}} \rightarrow \frac{\ket{\uparrow}_1\ket{\mathcal{S}^\dagger\vb n}_2 -i \ket{\downarrow}_1\ket{-\mathcal{S}^\dagger\vb n}_2}{\sqrt{2}} \otimes \ket{\uparrow}_{\text{dw}}, \;\; \text{and}\;\;  \ket{\uparrow}_1\otimes\ket{\uparrow}_2\otimes \ket{\uparrow}_{\text{dw}} \rightarrow \ket{\uparrow}_1\otimes\ket{-\mathcal{S}^\dagger\vb n}_2\otimes \ket{\uparrow}_{\text{dw}}. \)

\textbf{Case II.}|We now consider the initial state of the spin qubits to be $\ket{\uparrow}_1\otimes \ket{\downarrow}_2$. To entangle these two qubits, we initialize the domain wall in $\ket{\downarrow}_{\text{dw}}$. When the domain wall is approaching the second spin qubit, the state is $(\ket{\uparrow}_1 \ket{\vb n}_{\text{dw}} -i \ket{\downarrow}_1 \ket{-\vb n}_{\text{dw}}    )\otimes\ket{\downarrow}_2$, which evolves into the final entangled state $(\ket{\uparrow}_1\ket{\sigma_z\mathcal{S}^\dagger\vb n}_2-i\ket{\downarrow}_1\ket{-\sigma_z\mathcal{S}^\dagger\vb n}_2   )\otimes \ket{\downarrow}_{\text{dw}}$. On the other hand, when we flip the initial state of the domain wall to  $\ket{\uparrow}_{\text{dw}}$, the final state is  $\ket{\uparrow}_1\otimes \ket{-\sigma_z\mathcal{S}^\dagger\vb n}_2\otimes \ket{\downarrow}_{\text{dw}}$. We summarize as 
\( \ket{\uparrow}_1\otimes \ket{\downarrow}_2\otimes\ket{\downarrow}_{\text{dw}} \rightarrow \frac{\ket{\uparrow}_1\ket{\sigma_z\mathcal{S}^\dagger\vb n}_2-i\ket{\downarrow}_1\ket{-\sigma_z\mathcal{S}^\dagger\vb n}_2 }{\sqrt{2}}\otimes \ket{\downarrow}_{\text{dw}},\;\;\text{and}\;\; \ket{\uparrow}_1\otimes \ket{\downarrow}_2\otimes\ket{\uparrow}_{\text{dw}} \rightarrow \ket{\uparrow}_1\otimes \ket{-\sigma_z\mathcal{S}^\dagger\vb n}_2\otimes \ket{\downarrow}_{\text{dw}}. \)

\textbf{Case III.}|We consider the initial state of the spin qubits to be $\ket{\downarrow}_1\otimes \ket{\uparrow}_2$. Following the same procedure, we find that, under the same entanglement protocol, we have
\(\ket{\downarrow}_1\otimes \ket{\uparrow}_2 \otimes \ket{\uparrow}_{\text{dw}} \rightarrow \frac{\ket{\uparrow}_1\ket{\mathcal{S}^\dagger\vb n}_2 + i\ket{\downarrow}_1\ket{-\mathcal{S}^\dagger\vb n}_2}{\sqrt{2}}\otimes \ket{\uparrow}_{\text{dw}}, \;\; \text{and}\;\;  \ket{\downarrow}_1\otimes \ket{\uparrow}_2 \otimes \ket{\downarrow}_{\text{dw}} \rightarrow  \ket{\downarrow}_1\otimes\ket{\mathcal{S}^\dagger\vb n}_2 \otimes  \ket{\uparrow}_{\text{dw}}. \)

\textbf{Case IV.}|Similarly, for the initial state $\ket{\downarrow}_1\otimes \ket{\downarrow}_2$, we have the following map under the exactly the same entanglement protocol: 
\( \ket{\downarrow}_1\otimes \ket{\downarrow}_2\otimes\ket{\uparrow}_{\text{dw}}\rightarrow  \frac{ \ket{\uparrow}_1\ket{\sigma_z\mathcal{S}^\dagger\vb n}_2+i\ket{\downarrow}_1\ket{- \sigma_z \mathcal{S}^\dagger\vb n}_2  }{\sqrt{2}}\otimes \ket{\downarrow}_{\text{dw}}, \;\; \text{and}\;\; \ket{\downarrow}_1\otimes \ket{\downarrow}_2\otimes\ket{\downarrow}_{\text{dw}}\rightarrow   \ket{\downarrow}_1\otimes\ket{\sigma_z\mathcal{S}^\dagger\vb n}_2\otimes \ket{\downarrow}_{\text{dw}}.   \)

Therefore, we have demonstrated that the remote entanglement generation process can be turned on and off on demand by switching the DW initial state.

\subsection*{(V) Universality of the distant two-qubit logic gate}
\begin{figure}[hbt!]
\includegraphics[width=0.56\columnwidth]{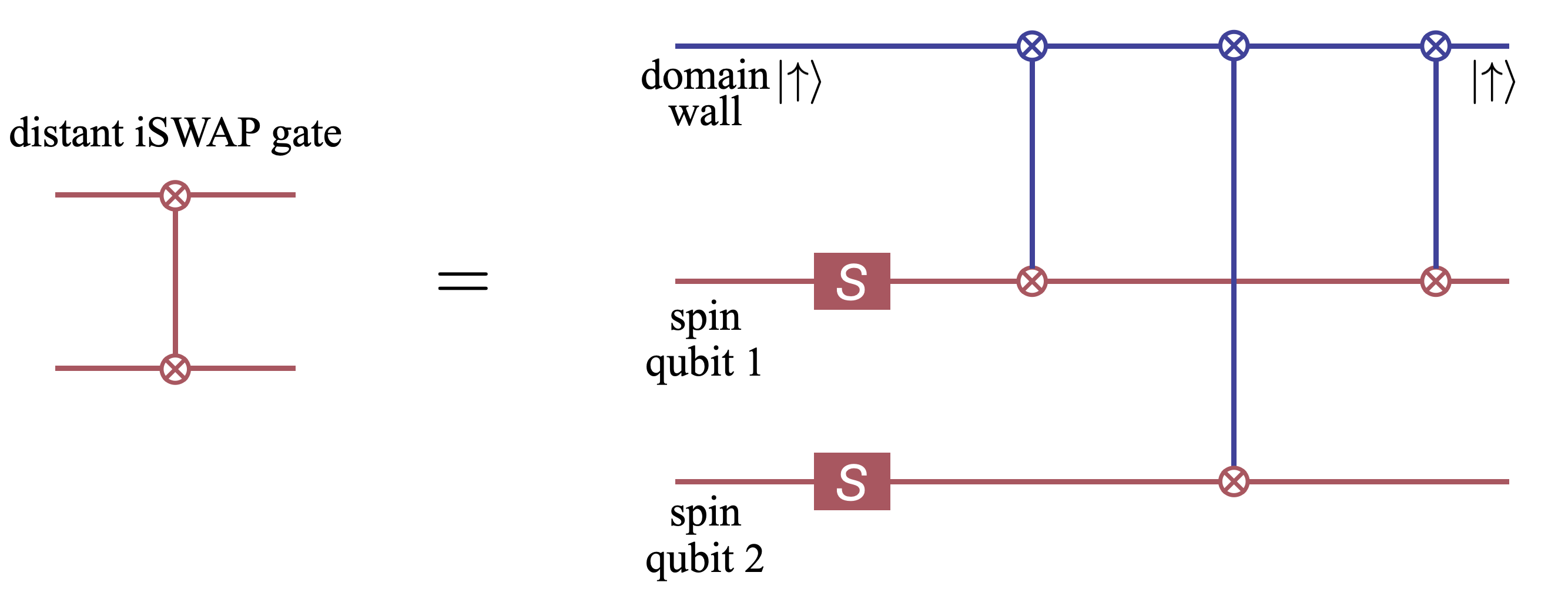}
\caption{Quantum circuit representation of a distant iSWAP gate. We first apply the phase gate $\mathcal{S}$ to the two spin qubits. Next, we move the domain wall (initially in state $\ket{\uparrow}$) across the first spin qubit to realize a local iSWAP gate, then across the second spin qubit to perform another iSWAP. Finally, we move the domain wall back to the first spin qubit, implementing another iSWAP, while disabling the interaction with the second spin qubit on the return path.} 
\label{sm1}
\end{figure}

In this section, we explicitly show that the mobile domain wall facilitates a distant iSWAP gate between two remote spin qubits, and combined with single-qubit rotations, enables universal quantum computation. The quantum circuit representation of such a remote two-qubit gate is shown in Fig.~\ref{sm1}. We assume the spin qubit 1 and qubit 2 are initially in  states: 
\( \ket{\psi}_1= a\ket{\uparrow}_1 +b\ket{\downarrow}_1, \;\;\; \text{and}\;\;\;  \ket{\phi}_2= c\ket{\uparrow}_2 +d\ket{\downarrow}_2,  \)
where $a, b, c, d$ are unknown complex numbers in general. We now apply phase gates $\mathcal{S}$ to these two qubits, resulting in the following states
\( \ket{\tilde{\psi}}_1= a\ket{\uparrow}_1 +i b\ket{\downarrow}_1, \;\;\; \text{and}\;\;\;  \ket{\tilde{\phi}}_2= c\ket{\uparrow}_2 +id\ket{\downarrow}_2.    \) 

We now move the domain wall at a proper velocity (initially in state $\ket{\uparrow}$) across the first spin qubit to realize a local iSWAP gate, which yields
\( \ket{\tilde{\psi}}_1 \otimes\ket{\uparrow}_{\text{dw}} \xrightarrow{\text{iSWAP}} \ket{\uparrow}_1\otimes (a \ket{\uparrow}_{\text{dw}} +b\ket{\downarrow}_{\text{dw}}  ).  \)
We continue moving the domain wall across the second spin qubit, leading to another local iSWAP gate:
\(  \ket{\tilde{\phi}}_2\otimes  (a \ket{\uparrow}_{\text{dw}} +b\ket{\downarrow}_{\text{dw}}  )  \xrightarrow{\text{iSWAP}}     ac\ket{\uparrow}_2\otimes\ket{\uparrow}_{\text{dw}} -ibc\ket{\downarrow}_2\otimes\ket{\uparrow}_{\text{dw}} +ad\ket{\uparrow}_2\otimes\ket{\downarrow}_{\text{dw}}+idb\ket{\downarrow}_2\otimes\ket{\downarrow}_{\text{dw}}.   \)
We now move the domain wall back to the first qubit. On its return path, we decouple it from the second spin qubit by shifting the qubit away from the racetrack using a local electric field. As the domain wall passes the first qubit again, it executes another local iSWAP gate:
\( \al{ &\ket{\uparrow}_1\otimes \left[  ac\ket{\uparrow}_2\otimes\ket{\uparrow}_{\text{dw}} -ibc\ket{\downarrow}_2\otimes\ket{\uparrow}_{\text{dw}} +ad\ket{\uparrow}_2\otimes\ket{\downarrow}_{\text{dw}}+idb\ket{\downarrow}_2\otimes\ket{\downarrow}_{\text{dw}} \right] \\
 &\xrightarrow{\text{iSWAP between the first qubit and DW}}    \ket{\uparrow}_{\text{dw}}\otimes \left[  ac\ket{\uparrow}_1\ket{\uparrow}_2 -ibc \ket{\uparrow}_1\ket{\downarrow}_2 -iad\ket{\downarrow}_1\ket{\uparrow}_2 +db\ket{\downarrow}_1\ket{\downarrow}_2  \right].      }\)
 Consequently, the domain wall returns to its initial state $\ket{\uparrow}_{\text{dw}}$, fully disentangled from the two spin qubits. Crucially, however, the spin qubit state undergoes a nontrivial evolution:
 \(   \ket{\text{initial}}=  \mqty(ac \\ ad\\ bc\\ bd)\rightarrow  \ket{\text{final}}= \mqty(ac \\ -ibc\\ -iad\\ bd) = \mqty(1 & & & \\ & 0 & -i & \\  & -i &  0 & \\ & & & 1  ) \ket{\text{initial}},  \)
which is equivalent to a remote iSWAP gate acting on the two distant qubits. 

We note that the general idea is to first transfer the state of the first spin qubit to the DW, then allow the DW to interact with the second spin qubit to realize an iSWAP gate. Finally, the DW returns to the first spin qubit to transfer its state back to the  qubit. This sequence effectively implements the iSWAP gate between the two spin qubits. A similar approach can be applied to realize a $\sqrt{\text{iSWAP}}$ gate. Instead of performing an iSWAP gate in the second step, we can implement a $\sqrt{\text{iSWAP}}$ gate between the second spin qubit and the DW. Then, this process would ultimately result in a $\sqrt{\text{iSWAP}}$ gate between the distant qubits.

\subsection*{(VI) Detailed derivations of protocol fidelity}
In this section, we present a detailed analysis of the fidelity of the entanglement protocol and derive Equations (6) and (7) in the main text. We consider both coherent  and incoherent errors. Calculating the  fidelity of a quantum operation is essential to evaluate how closely a real operation $\mathcal{U}_R$  matches an ideal one $\mathcal{U}_T$, which can be represented as 
\(  \mathcal{F}= \frac{1}{\dim[\mathcal{U}]} \Tr\big[  \mathcal{U}_R  \mathcal{U}_T^\dagger \big].  \label{fidelity}  \)

\textbf{Coherent errors.}|In our previous analysis, we have dropped the effective magnetic field $h_z(\mathcal{X})$ acting on the spin qubit due to its coupling to the racetrack. Taking this term into account, the total Hamiltonian describing the interaction between spin qubit and domain wall qubit is given by
\( \al{\mathcal{H}(t)&= - \frac{\hbar \omega_s}{2} \sigma_y -h_z[\mathcal{X}(t)] \sigma_z - \frac{t_g}{2} \tau_x -\mathcal{J} g[\mathcal{X}(t)] \sigma_x \otimes \tau_z  \\
   &=- \frac{\hbar \omega_s}{2} \hat{\sigma}_z + h_z[\mathcal{X}(t)] \hat{\sigma}_y   - \frac{t_g}{2} \hat{\tau}_z +\mathcal{J} g[\mathcal{X}(t)] \hat{\sigma}_x \otimes \hat{\tau}_x,  }   \)
where we again rotate the spin axis for convenience.
We now conduct a time-dependent spin rotation around $x$ axis in the spin space:
\( U[\theta(t)]= \exp{i \left[\theta(t)-\frac{\pi}{2}\right]\frac{\hat{\sigma}_x}{2}}, \;\; \text{with}\;\; \cos\theta(t)= \frac{2h_z(t)/\hbar}{\sqrt{\omega_s^2+(2h_z/\hbar)^2}} \rightarrow \theta(t)\approx \frac{\pi}{2}- \frac{2h_z(t)}{\hbar\omega_s}, \)
where we have used the fact that the induced effective magnetic field is one order of magnitude smaller than the qubit frequency $h_z \sim \mathcal{J} \ll \hbar\omega_s$. We note that the spin axis is tilted by a small angle ${2h_z(t)}/{\hbar\omega_s}$ due to this effective magnetic field. Then the Hamiltonian is given by: 
\( \al{  \mathcal{H}_{R} &=U^\dagger[\theta(t)] H U[\theta(t)] + i\hbar (\partial_t U^\dagger[\theta(t)]) U[\theta(t)] =- \frac{\hbar \sqrt{ \omega_s^2+ [2h_z(t)/\hbar]^2  }}{2} \hat{\sigma}_z + \frac{\hbar\dot{\theta}(t)}{2} \hat{\sigma}_x - \frac{{t}_g}{2} \hat{\tau}_z + \mathcal{J}(t) \hat{\sigma}_x\otimes \hat{\tau}_x  \\
 &\approx -\frac{\hbar \omega_s}{2} \hat{\sigma}_z  - \frac{{t}_g}{2} \hat{\tau}_z + \mathcal{J}(t) \hat{\sigma}_x\otimes \hat{\tau}_x. }  \)
Here, we have neglected two terms. The first term is the qubit frequency renormalization due to the effective magnetic field, $\delta\omega_s=\sqrt{\omega_s^2+(2h_z/\hbar)^2}-\omega_s$. We will return to this frequency shift later.  The second term we dropped is rooted in the time dependence of the tilting angle $\dot{\theta}(t) \approx \Delta\theta/T_g \sim 2\pi \times 2\, \text{MHz}$, where $\Delta \theta \sim 0.03$ and $T_g \sim 2\, \text{ns}$ is the local two-qubit gate time. The error induced by this term would be negligible compared to the error that we are considering here induced by the effective magnetic field, which is on the order of $100\, \text{MHz}$. Thus, we will neglect this term in the following analysis. 

We now switch to the interaction picture by the unitary transformation $\mathcal{U}(t)=\exp{i\omega_s t( \hat{\sigma}_z+\hat{\tau}_z )/2}. $ Under the rotating wave approximation and working at $t_g=\hbar \omega_s$, we arrive at the Hamiltonian $H_I(t)\approx \mathcal{J}[g(t)/2]( \hat{\sigma}_x \otimes \hat{\tau}_x + \hat{\sigma}_y \otimes \hat{\tau}_y)$. Thus, the real operation in the original frame is given by
\( \mathcal{U}_R= {U}[\theta(T_g)] \mathcal{U}(T_g) \exp{-\frac{i}{\hbar}\int^{T_g}_{0} d\tau \, H_I(\tau) } \mathcal{U}^\dagger(0) U^\dagger[\theta(0)].  \)
On the other hand, the target (ideal) evolution is given by
\(  \mathcal{U}_T=  \mathcal{U}(T_g) \exp{-\frac{i}{\hbar}\int_0^{T_g} d\tau \, H_I(\tau) } \mathcal{U}^\dagger(0), \)
where  $T_g$ is the gate operational time.  

{ We can now evaluate: 
 \(     \mathcal{U}_R  \mathcal{U}_T^\dagger= {U}[\theta(T_g)] \mathcal{U}(T_g) \exp{-\frac{i}{\hbar}\int^{T_g}_{0} d\tau \, H_I(\tau) }  U^\dagger[\theta(0)]   \exp{\frac{i}{\hbar}\int_0^{T_g} d\tau H_I(\tau) } \mathcal{U}^\dagger(T_g).   \)  
 We first note that: 
 \(  \mathcal{U}(T_g) \exp{-\frac{i}{\hbar}\int^{T_g}_{0} d\tau \, H_I(\tau) } = \mqty[ \exp{i\omega_s T_g} & 0 & 0 & 0 \\ 0 & \cos\Phi  & -i\sin\Phi & 0 \\ 0 & -i\sin\Phi & \cos \Phi & 0 \\ 0& 0& 0& \exp{-i\omega_s T_g}   ].   \)
   We also know that: 
   \(   U[\theta(0)]=\exp{-i\delta\theta \hat{\sigma}_x/2}, \;\;\;\text{and}\;\;\;  U[\theta(T_g)]=\exp{i\delta\theta \hat{\sigma}_x/2},    \)     
   where $\delta\theta= 2h_z(0)/(\hbar \omega_s)=4\mathcal{J}/(\hbar \omega_s)$ and we have used $h_z(0)=-h_z(T_g)=2\mathcal{J}$. 
From these analysis, we can immediately obtain the expression for the fidelity:}
\( \mathcal{F}=\cos^2\frac{\delta\theta}{2}- \cos\Phi \cos(\omega_sT_g)\sin^2\frac{\delta\theta}{2}.  \)
Here,  $\Phi=[\mathcal{J}/(\hbar v)]\int g(\mathcal{X})d\mathcal{X}$ and $\delta \theta$ is the maximal tilting angle of the spin axis due to the effective magnetic field $h_z$. Then the error induced by this spin axis tilting is 
\( 1-\mathcal{F}=\sin^2\frac{\delta\theta}{2} +\cos\Phi\cos(\omega_s T_g) \sin^2\frac{\delta\theta}{2} \leq 2\sin^2 \frac{\delta\theta}{2}.  \label{fidelity1}  \)

In the discussion above, we have pointed out that the effective magnetic field also shifts the qubit frequency. Besides, in experiments, it may also be difficult to make the two qubits exactly on resonance $\hbar\omega_s=t_g$. So here we investigate the two-qubit gate error induced by a detuning between the two frequencies. We consider the following Hamiltonian: 
\(\mathcal{H}(t)=-\frac{\hbar(\omega_s+\delta \omega_s)}{2} \hat{\sigma}_z - \frac{\hbar\omega_s}{2} \hat\tau_z +\mathcal{J}g(t)\hat\sigma_x\otimes \hat\tau_x. \)
We again move to the interaction picture and apply the rotating wave approximation, yielding
\(  \mathcal{H}_I(t)\approx \mathcal{J}g(t)\left[ \hat\sigma_+ \hat\tau_- e^{-i\delta \omega_s t} +\text{h.c.}  \right] = \frac{\mathcal{J}g(t)\cos(\delta \omega_s t)}{2} (\hat\sigma_x\hat\tau_x+\hat\sigma_y \hat\tau_y) - \frac{\mathcal{J}g(t) \sin(\delta\omega_s t)}{2} (\hat\sigma_x \hat\tau_y - \hat\sigma_y \hat\tau_x).  \)
For simplicity, we consider $\delta\omega_s$ to be on the order of $20\, \text{MHz}$ or less (the frequency shift due to the effective magnetic field $h_z$ is about $2\, \text{MHz}$), which implies $\delta\omega_s T_g\ll 1$. We then approximate the Hamiltonian with
\(   \mathcal{H}_I(t) \approx\frac{\mathcal{J}g(t)}{2} (\hat\sigma_x\hat\tau_x+\hat\sigma_y \hat\tau_y) - \frac{\mathcal{J}g(t) \sin(\delta\omega_s t)}{2} (\hat\sigma_x \hat\tau_y - \hat\sigma_y \hat\tau_x).  \)
To see the effect of the last term due to finite $\delta\omega_s$, we apply the unitary transformation
\(  \mathcal{U}_2(t)=\exp{-\frac{i}{2}\Phi(t) (\hat\sigma_x\hat\tau_x+\hat\sigma_y \hat\tau_y) }, \;\;\text{with}\;\;\Phi(t)=(\mathcal{J}/\hbar)\int^t_0g(\tau) d\tau,  \)
which leads to the following Hamiltonian
\(  \mathcal{H}_2(t)= - \frac{\mathcal{J}g(t) \sin(\delta\omega_s t)}{2} \mathcal{U}_2^\dagger(t) (\hat\sigma_x \hat\tau_y - \hat\sigma_y \hat\tau_x) \mathcal{U}_2(t)= - \frac{\mathcal{J}g(t) \sin(\delta\omega_s t)}{2}  \left[ \cos ({2}\Phi(t)) (\hat\sigma_x \hat\tau_y - \hat\sigma_y \hat\tau_x) +\sin({2}\Phi(t)) (\hat\sigma_z- \hat\tau_z)  \right].  \)
Then the fidelity is given by 
\(  \mathcal{F}=\frac{1}{4}\Tr \mathcal{T}\exp{-\frac{i}{\hbar}\int^{T_g}_0\!\!d\tau\, \mathcal{H}_2(\tau) } \approx 1- \bigg[\int^{T_g}_0 d\tau \frac{\mathcal{J}g(\tau) \sin(\delta\omega_s \tau) \cos {2} \Phi(\tau) }{2\hbar}\bigg]^2 - \bigg[\int^{T_g}_0 d\tau \frac{\mathcal{J}g(\tau) \sin(\delta\omega_s \tau) \sin {2} \Phi(\tau) }{2\hbar}\bigg]^2. \)
Thus the corresponding error is 
\( \al{ 1-\mathcal{F}&=  \bigg[\int^{T_g}_0 d\tau \frac{\mathcal{J}g(\tau) \sin(\delta\omega_s \tau) \cos{2}\Phi(\tau) }{2\hbar}\bigg]^2 + \bigg[\int^{T_g}_0 d\tau \frac{\mathcal{J}g(\tau) \sin(\delta\omega_s \tau) \sin{2}\Phi(\tau) }{2\hbar}\bigg]^2 \\
&= \frac{\mathcal{J}^2}{4\hbar^2}\int^{T_g}_0d\tau \int^{T_g}_0ds\, g(\tau) g(s) \sin(\delta\omega_s \tau)\sin(\delta\omega_s s) [ \cos{2}\Phi(\tau) \cos {2}\Phi(s) +\sin {2}\Phi(\tau)\sin {2}\Phi(s)    ] \\
&= \int^{T_g}_0 \frac{dt}{2\hbar} \int^{T_g}_0  \frac{ds}{2\hbar} \check{\mathcal{J}}(t)\check{\mathcal{J}}(s) \cos[{2}(\Phi(t)-\Phi(s))],\;\;\text{with} \;\; \check{\mathcal{J}}(t)=\mathcal{J}g(t)\sin(\delta\omega_s t).
   }\)
 Combing with Eq.~\eqref{fidelity1}, we have derived the Equation (6) in the main text.   
 
\textbf{Error due to noise.}|Here we discuss the error induced by the noise in the system. We note that while the spin qubits can reach 
decoherence times of up to 10$\mu$s in experiments, the domain wall qubit lifetime is estimated to be of submicrosecond ($\sim 0.5\,\mu$s) with feasible experimental parameters in current materials. Consequently, the fidelity of our system is primarily limited by the noise affecting the domain wall qubit.
The dominant fluctuating term in the domain wall qubit Hamiltonian arises from the fluctuations of the azimuthal angle of the domain wall texture in spin space, represented as $\delta H=[h(t)/2]\tau_z$. Here, $h(t)$ is the fluctuating field, characterized by the classical ensemble average and their correlation function: $\langle h(t) \rangle=0$ and $\langle h(t) h(s)\rangle=S(t-s)$, where $S(t)$ is related to the dissipation parameter via the fluctuation dissipation theorem and the corresponding spectral function is $S(\omega)=\hbar^2 \alpha N \omega \coth[ \hbar \omega/(2k_BT_0) ]$. Here $\alpha$ is Gilbert damping and $T_0$ is the temperature. Let us write down the  Hamiltonian for the domain wall: 
\( \mathcal{H}(t)=-\frac{t_g}{2}\hat\tau_z - \frac{\hbar \delta v[\mathcal{X}(t)]}{\ell_{\text{so}}}\hat\tau_x + \frac{h(t)}{2}\hat\tau_x,  
 \)
where we consider the domain wall velocity profile on the racetrack to be $v(\mathcal{X})=v_0+\delta v[\mathcal{X}(t)]$ and we also have rotated the spin axis. We again rotate the spin axis such that
\( \mathcal{H}(t)=- \frac{\hbar \Omega(t)}{2} \hat{\tau}_z + \frac{h(t)}{2} \left\{ \sin[\theta(t)] \hat{\tau}_x  -\cos[\theta(t)] \hat{\tau}_z \right\},\;\text{with}\;\frac{\hbar\Omega(t)}{2}=\sqrt{(t_g/2)^2 +[\hbar \delta v(t)/\ell_{\text{so}}]^2 } \;\text{and}\; \tan\theta(t)=-\frac{t_g\ell_{\text{so}}}{2\hbar \delta v(t)}. \)
We now apply the following unitary transformation: 
\( \mathcal{U}_1= \exp{i\frac{\tilde\Phi(t)}{2}\hat{\tau}_z}, \;\;\text{with}\;\; \tilde\Phi(t)=\int^t_0d\tau\, \Omega(\tau), \)
which yields 
\( \mathcal{H}_1(t)=\frac{h(t)}{2} \sin\theta(t)\, \left[ \cos\tilde\Phi(t)\hat{\tau}_x + \sin\tilde\Phi(t)\hat{\tau}_y  \right] - \frac{h(t)}{2} \cos\theta(t)\, \hat{\tau}_z.  \)
The fidelity therefore is given by
\(\al{ \mathcal{F} &= \Tr\mathcal{T}\exp{-\frac{i}{\hbar} \int^T_0d\tau\, \mathcal{H}_1(\tau) } = 1- \frac{1}{2\hbar^2} \int^T_0 d\tau \int^T_0 d\tau^\prime \Tr \mathcal{T}[\mathcal{H}_1(\tau) \mathcal{H}_1(\tau^\prime)]\\
&=1-\frac{1}{8\hbar^2} \int^T_0 d\tau \int^T_0 d\tau^\prime h(\tau) h(\tau^\prime) \sin\theta(\tau) \, \sin\theta(\tau^\prime) \cos[\tilde\Phi(\tau)-\tilde\Phi(\tau^\prime)] - \frac{1}{8\hbar^2} \int^T_0 d\tau \int^T_0 d\tau^\prime h(\tau) h(\tau^\prime) \cos\theta(\tau) \, \cos\theta(\tau^\prime) \\
&=1- \frac{1}{8\hbar^2} \int^\infty_{-\infty}\frac{d\omega}{2\pi} S(\omega) F(\omega, T),   }\)
where $T$ is the total shuttling time of the domain wall and $F(\omega, T)$ is the effective filter function determined by:
\(  F(\omega, T)= \Big| \int^T_0 d\tau\, e^{-i\omega \tau} \sin\theta(\tau) \cos\tilde\Phi(\tau)  \Big|^2 + \Big| \int^T_0 d\tau\, e^{-i\omega \tau} \sin\theta(\tau) \sin\tilde\Phi(\tau)  \Big|^2 +  \Big| \int^T_0 d\tau\, e^{-i\omega \tau} \cos\theta(\tau)  \Big|^2.   \)
For convenience, we can introduce a  time-dependent vector:
\(  \vb m(\tau)=\big[ \sin\theta(\tau) \cos\tilde\Phi(\tau), \sin\theta(\tau) \sin\tilde\Phi(\tau),  \cos\theta(\tau)  \big].  \)
The filter function then can be written in the following compact form: 
\(  F(\omega, T)= |\vb*{\mathcal{M}}(\omega, T)|^2, \;\;\text{with}\;\;  \vb*{\mathcal{M}}(\omega, T) =\int^T_0d\tau\, e^{-i\omega \tau} \vb m(\tau).  \) 
This gives us the Equation (7) in the main text. In a special case,  this result is greatly simplified when the velocity variation is a constant $\delta v$ and thus $\Omega(t)$ is also a constant denoted as $\Omega$. The error or infidelity can be reduced to
\( 1-\mathcal{F}= \frac{T}{4} \frac{{\cos}^2\theta}{2\hbar^2}S(0) + \frac{T}{4} \frac{{\sin}^2\theta}{4\hbar^2}[S(\Omega)+S(-\Omega)]=\frac{T}{4T_2},  \)
 with $1/T_2=1/T_\varphi+1/(2T_1)$ where $1/T_\varphi={\cos}^2\theta S(0)/(2\hbar^2)$ and $1/T_1=[{\sin}^2\theta/(2\hbar^2)][S(\Omega)+S(-\Omega)]$.

%